\documentclass[aps,preprint,showpacs,showkeys]{revtex4}

\begin{document}


\newcommand\beq{\begin{equation}}
\newcommand\eeq{\end{equation}}
\newcommand\beqa{\begin{eqnarray}}
\newcommand\eeqa{\end{eqnarray}}
\newcommand\ket[1]{|#1\rangle}
\newcommand\bra[1]{\langle#1|}
\newcommand\scalar[2]{\langle#1|#2\rangle}

\newcommand\jo[3]{\textbf{#1}, #3 (#2)}


\title{\Large\textbf{On the question of secret probability distributions in quantum bit commitment}}

\author{Chi-Yee Cheung}

\email{cheung@phys.sinica.edu.tw}

\affiliation{Institute of Physics, Academia Sinica\\
             Taipei, Taiwan 11529, Republic of China\\}


\begin{abstract}
The proof of the No-Go Theorem of unconditionally secure quantum bit commitment depends on the assumption that Alice knows every detail of the protocol, including the probability distributions associated with all the random variables generated by Bob. We argue that this condition may not be universally satisfied. In fact it can be shown that when Bob is allowed to use a secret probability distribution, the joint quantum state is inevitably mixed. It is then natural to ask if Alice can still cheat.
A positive answer has been given by us \cite{Cheung07} for the perfect concealing case. In this paper, we present a simplified proof of our previous result, and extend it to cover the imperfect concealing case as well.
\end{abstract}



\maketitle


Quantum bit commitment is an important two-party primitive in quantum cryptography, because a secure quantum bit commitment protocol can be used to guarantee the security of a number of other cryptographic protocols.\cite{Brassard96,Blum83,Bennett91,Crepeau94,Yao95,Mayers96,Brassard88,Kilian88,Crepeau95}

Bit commitment involves a sender (Alice) and a receiver (Bob). Alice commits to Bob a secret bit $b\in\{0,1\}$ and at the same time provides him with a piece of evidence. When Alice unveils the secret bit sometime in the future, Bob can check the evidence and verify that the unveiled bit is the same as what was committed by Alice in the beginning. Now Alice and Bob do not trust each other, in the sense that Bob would try to gain information about the committed bit (from the provided evidence) before Alice unveils it, and Alice would try to change her commitment if it is to her advantage to do so. A bit commitment protocol is said to be secure if, (1) Bob cannot know the value of $b$ before Alice reveals it (concealing), and (2) Alice cannot change $b$ without Bob's knowledge (binding).

In quantum bit commitment (QBC), Alice and Bob together execute a series of quantum and classical operations during the commitment procedure, such that in the end Bob holds a quantum state $\rho_B^b$ which serves as the evidence of Alice's commitment. If
\beq
\rho_B^0=\rho_B^1,
\eeq
the protocol is said to be perfect concealing,
and obviously Bob is not able to extract any information about the value of $b$ from the $\rho^b_B$ in his possession. For imperfect protocols, the two density matrices are equal only asymptotically as the security parameter
$N\rightarrow\infty$. For large but finite $N$, one has
\beq
\rho^0_B\approx\rho^1_B,
\eeq
so that Bob's knowledge of $b$ (before Alice unveils it) vanishes in the limit $N\rightarrow\infty$.

If a QBC protocol is secure even if both Alice and Bob had unlimited computing power, then it is said to be unconditionally secure. Unfortunately unconditionally secure quantum bit commitment is ruled out by a no-go theorem \cite{Mayers97,LoChau97}. In essence the theorem says that, if a protocol is concealing to Bob, then it is cannot be binding to Alice. That means, if $\rho_B^0=\rho_B^1$, then using a unitary transformation $U_A$ Alice has the freedom to rotate $\rho_B^0$ into $\rho_B^1$ or vice versa by operating on her own quantum particles only. As a result she can commit to one bit value and safely unveils another without Bob's knowledge. It is not hard to see that this no-go conclusion depends on the assumption that Alice can always calculate $U_A$ without the help of Bob, which is equivalent to saying that she knows ``every detail of the protocol, including the distribution of probability of a random variable generated by another participant" \cite{Brassard97}. However it is not obvious that this condition is universally valid in all possible QBC protocols. And when it is not, the validity of the no-go proof needs to be reexamined.

In the picture where all random variables are purified (that is, where all unrevealed classical choices are left undetermined by quantum entanglement), the only parameters that can remain secret are probability distributions. The problem of secret probability distributions in QBC has been addressed partially in \cite{Cheung07}, where we showed that for perfect concealing protocols Alice can still safely cheat even if she does not know the probability distribution Bob used to entangle a random variable. The purpose of this paper is to provide a simplified proof of our earlier result, furthermore we show that the same conclusion applies to imperfect concealing protocols as well.

To facilitate our discussion, we shall first outline the proof of the no-go theorem below. The crucial observation is that, using quantum entanglement, Alice and Bob can keep all undisclosed classical information undetermined and stored at the quantum level. In other words, they can always choose to delay any prescribed classical actions without consequences until it is required to disclose the outcomes. Then one can assume that, at the end of the
commitment procedure, there exists a pure state
$\ket{\psi^b_{AB}}$ in the joint Hilbert space
of Alice and Bob $H_A\otimes H_B$. $\ket{\psi^b_{AB}}$
is called a purification of the quantum state
$\rho^b_B$ in Bob's hand, such that
\beq
{\rm Tr}_A~\ket{\psi^b_{AB}}\bra{\psi^b_{AB}}
=\rho_B^b.
\eeq
Note that, because $H_A$ and $H_B$ are disjoint, whether Bob actually purifies or not is irrelevant to Alice, without loss of generality she can assume he always does.
In general, purification requires access to fully functioning
quantum computers, which is nevertheless not a problem since both participants are assumed to have unlimited computational power.

For the perfect concealing case, where $\rho^0_B=\rho^1_B$,
it can be shown that the two purifications $\ket{\psi^0_{AB}}$ and $\ket{\psi^1_{AB}}$ are related by a unitary transformation on Alice's side \cite{Hughston93}, namely,
\beq
\ket{\psi^1_{AB}}=U_A\ket{\psi^0_{AB}}
\eeq
If Alice knows all the parameters used by Bob, then she can compute and then execute $U_A$ without Bob's help. That means she can commit to $b=0$ but safely unveil $b=1$ (or vice versa); this is called the entanglement attack. It follows that perfect concealing protocols are not binding.

Next we consider the imperfect case where $\rho^0_B$ and $\rho^1_B$ are close but unequal. Quantitatively that means the fidelity $F$ of the two density matrices is close to one. Using Uhlmann's theorem \cite{Jozsa94}, we can write
\beq
F(\rho^0_B,\rho^1_B)={\rm max}\,
|\scalar{\phi^0_{AB}}{\phi^1_{AB}}|=1-\epsilon,
\label{imperfect}
\eeq
where $\epsilon\rightarrow 0$ as the security parameter $N\rightarrow\infty$, $\ket{\phi^0_{AB}}$ and $\ket{\phi^1_{AB}}$ are purifications of $\rho^0_B$ and $\rho^1_B$ respectively, and the maximization is over all possible purifications. Uhlmann's theorem also implies that, for a fixed purification $\ket{\psi^1_{AB}}$ of $\rho^1_B$, there exists an optimal purification $\ket{\varphi^0_{AB}}$ of $\rho^0_B$ such that
\beq
F(\rho^0_B,\rho^1_B)=|\scalar{\psi^1_{AB}}{\varphi^0_{AB}}|
=1-\epsilon.
\label{fidelity-1}
\eeq
Since $\ket{\psi^0_{AB}}$ and $\ket{\varphi^0_{AB}}$ are two purifications of the same density matrix $\rho^0_B$, by the previous argument, there must exist a unitary transformation $U_A$ such that $\ket{\varphi^0_{AB}}=U_A\ket{\psi^0_{AB}}$, so
\beq
|\bra{\psi^1_{AB}}U_A\ket{\psi^0_{AB}}|=1-\epsilon.
\label{fidelity-2}
\eeq
So Alice can also cheat when $\rho^0_B\approx\rho^1_B$, provided that she knows $U_A$.

Mathematically the no-go theorem only proves that there exists a unitary transformation $U_A$ which can turn $\psi^0_{AB}$ to $\psi^1_{AB}$, either exactly or asymptotically. As mentioned before, for the no-go theorem to be valid, one must also assume that Alice knows how to calculate $U_A$ by herself in every possible protocol. But that is by no mean obvious.

For example it may occur that the wavefunction $\ket{\psi^b_{AB}}$ depends on a certain parameter $\omega$ secretly chosen by Bob, then $U_A$ may also depend on $\omega$ and it would be unknown to Alice, unless proven otherwise. If so, could Alice still cheat?

One may doubt if this is a valid question, for what we are saying is that $\ket{\psi^b_{AB}}$ may be unknown to Alice and she is actually dealing with a mixed state, while as we saw the proof of the no-go theorem depends critically on the assumption that $\ket{\psi^b_{AB}}$ is pure. The original idea of the no-go proof is that whenever there is a random variable which renders the quantum state a mixed one, Alice can always work with the corresponding purified state. But that is possible only if she knew the probability distribution associated with the random variable in question. However if the probability distribution ($\omega$) is unknown, then the state is inevitably a mixed one, and any further purification attempt using another unknown probability distribution will not change that.

So the question being raised here is this: If a protocol allows Bob to choose a probability distribution $\omega$ which is not disclosed to Alice, could she still cheat by entanglement attack?
Unfortunately the answer is positive for both perfect and imperfect concealing protocols, as we shall show in the following.

Consider first the perfect concealing case. It has been discussed in \cite{Cheung07}, and we are presenting here a simplified and improved proof. Suppose $\omega_1$ and $\omega_2$ are any two possible probability distributions that Bob can use, the concealing condition implies that
\beqa
&&\ket{\psi^1_{AB}(\omega_1)}
=U_A(\omega_1)\ket{\psi^0_{AB}(\omega_1)},\label{UA1}\\
&&\ket{\psi^1_{AB}(\omega_2)}
=U_A(\omega_2)\ket{\psi^0_{AB}(\omega_2)},\label{UA2}
\eeqa
where $U_A(\omega_1)$ and $U_A(\omega_2)$ are unitary operators acting on Alice's particles.
Obviously Bob has the freedom to entangle his choices, in which case the overall state is given by
\beq
\ket{\Psi_{AB}^b}=\sqrt{p}\,\ket{\psi^b_{AB}(\omega_1)}\ket{\lambda_1}
             +\sqrt{1-p}\,\ket{\psi^b_{AB}(\omega_2)}\ket{\lambda_2},
             \label{entangle}
\eeq
where $p$ is a real number, $0<p<1$,  $\ket{\lambda_{1,2}}$ are ancilla states controlled by Bob, and $\scalar{\lambda_1}{\lambda_2}=0$.
The protocol should remain concealing, so $\ket{\Psi^0_{AB}}$ and $\ket{\Psi^1_{AB}}$ are again connected by a unitary transformation:
\beq
\ket{\Psi^1_{AB}}=\tilde U_A\ket{\Psi^0_{AB}},
\eeq
where $\tilde U_A$ may or may not depend on $p$, $\omega_1$, and $\omega_2$.
Since the ancilla states $\ket{\lambda_1}$ and $\ket{\lambda_2}$ are not affected by $U_A$, and they are orthogonal, it is easy to see that
\beqa
&&\ket{\psi^1_{AB}(\omega_1)}=\tilde U_A\ket{\psi^0_{AB}(\omega_1)},\\
&&\ket{\psi^1_{AB}(\omega_2)}=\tilde U_A\ket{\psi^0_{AB}(\omega_2)}.
\eeqa
Comparing these relations with with Eqs. (\ref{UA1},\ref{UA2}), we get
\beq
U_A(\omega_1)=\tilde U_A=U_A(\omega_2),
\eeq
for arbitrary $\omega_1$ and $\omega_2$. Hence $\tilde U_A$ depends neither on $p$ nor $\omega$. Therefore, as long as $\rho_B^0=\rho_B^1$, Alice can calculate $U_A$ without the knowledge of the $\omega$ actually employed by Bob.

Next we consider the imperfect case, where the density matrices on Bob's side $\rho_B^0$ and $\rho_B^1$ are close but not equal.
As in the perfect concealing case, if $\omega_1$ and $\omega_2$ are two possible choices for Bob, then the concealing condition guarantees that there exist optimal unitary operators $U_A(\omega_1)$ and $U_A(\omega_2)$ such that,
\beqa
|\bra{\psi^1_{AB}(\omega_1)}U_A(\omega_1)
\ket{\psi^0_{AB}(\omega_1)}|&=&1-\epsilon_1,\label{abs1}\\
|\bra{\psi^1_{AB}(\omega_2)}U_A(\omega_2)
\ket{\psi^0_{AB}(\omega_2)}|&=&1-\epsilon_2\label{abs2},
\eeqa
where $0 < \epsilon_{1,2} < 1$, and $\epsilon_{1,2}\rightarrow 0$ as the security parameter $N$ approaches infinity. As before when Bob entangles his choices as in Eq. (\ref{entangle}), there exists a $\tilde U_A$ such that
\beq
|\bra{\Psi^1_{AB}}\tilde U_A\ket{\Psi^0_{AB}}|=1-\tilde\epsilon,\label{teps}
\eeq
where $\tilde\epsilon\rightarrow 0$ as $N\rightarrow\infty$. Substituting Eq. (\ref{entangle}) into this equation gives
\beqa
&&p|\bra{\psi^1_{AB}(\omega_1)}\tilde U_A
\ket{\psi^0_{AB}(\omega_1)}|\nonumber\\
&&\quad+(1-p)|\bra{\psi^1_{AB}(\omega_2)}\tilde U_A
\ket{\psi^0_{AB}(\omega_2)}|\ge 1-\tilde\epsilon.\qquad\label{expand}
\eeqa
Let
\beqa
|\bra{\psi^1_{AB}(\omega_1)}\tilde U_A\ket{\psi^0_{AB}(\omega_1)}|&=&1-\delta_1,\label{eps1}\\
|\bra{\psi^1_{AB}(\omega_2)}\tilde
U_A\ket{\psi^0_{AB}(\omega_2)}|&=&1-\delta_2,\label{eps2}
\eeqa
where $0\le\delta_{1,2}\le1$. Then Eq. (\ref{expand}) gives
\beq
\tilde\epsilon\,\ge\, p\delta_1+(1-p)\delta_2\, >\, 0, \label{epsdelta}
\eeq
for arbitrary $p$, which implies that as the security parameter $N\rightarrow\infty$,
\beqa
&&\delta_1\rightarrow 0,\\
&&\delta_2\rightarrow 0,
\eeqa
like $\tilde\epsilon$.

Comparing Eq. (\ref{eps1}) and Eq. (\ref{eps2}) with
Eq. (\ref{abs1}) and Eq. (\ref{abs2}), respectively, we get
\beq
U_A(\omega_1)\approx\tilde U_A\approx U_A(\omega_2),
\eeq
for arbitrary $\omega_1$ and $\omega_2$, such that in the limit of  $N\rightarrow\infty$,
\beq
U_A(\omega_1)=\tilde U_A=U_A(\omega_2).
\eeq
Consequently Alice only needs to calculate $U_A(\omega)$ for any value of $\omega$, and she can use it to change her committed bit if she prefers - her chance of being discovered approaches zero when the security parameter $N$ approaches infinity.

Conversely, it is easy to see that if in any protocol one finds that the operator $U_A$ between $\psi^0_{AB}(\omega)$ and $\psi^1_{AB}(\omega)$ depends on $\omega$, then this protocol cannot be concealing, because when Bob entangles as in Eq. (\ref{entangle}), the resulting pure states $\ket{\Psi^0_{AB}}$ and $\ket{\Psi^1_{AB}}$ are not connected by an unitary transformation operating in Alice's Hilbert space, implying that $\tilde\rho_B^0\ne\tilde\rho_B^1.$

In summary we have argued in this paper that Alice cannot possibly know all the probability distributions used by Bob, because they do not trust each other. Then for a complete proof of the
no-go result,
one must also address the following question: In protocols where Bob is allowed to use probability distributions unknown to Alice during the commitment phase, can Alice still apply the entanglement attack? The answer we have arrived at is positive for both perfect and imperfect concealing cases, so unconditionally secure quantum bit commitment remains impossible.




\end{document}